\documentclass[apl,showpacs,twocolumn,preprintnumbers]{revtex4}
\usepackage{graphicx}

\newcommand{\fref}[1]{Fig.~\ref{#1}}

\begin{document}
\title{Shear effects in lateral piezoresponse force microscopy at 180$^\circ$ ferroelectric domain walls}
\author{J. Guyonnet}
\email{jill.guyonnet@unige.ch}
\affiliation{D\'epartement de Physique de la Mati\`ere Condens\'ee, Universit\'e de Gen\`eve, 24 Quai Ernest Ansermet, 1211 Geneva 4, Switzerland}
\author{H. B\'ea}
\affiliation{D\'epartement de Physique de la Mati\`ere Condens\'ee, Universit\'e de Gen\`eve, 24 Quai Ernest Ansermet, 1211 Geneva 4, Switzerland}
\author{F. Guy}
\affiliation{hepia, 4 rue de la Prairie, 1202 Geneva, Switzerland}
\author{S. Gariglio}
\affiliation{D\'epartement de Physique de la Mati\`ere Condens\'ee, Universit\'e de Gen\`eve, 24 Quai Ernest Ansermet, 1211 Geneva 4, Switzerland}
\altaffiliation{hepia, 4 rue de la Prairie, 1202 Geneva, Switzerland}
\author{S. Fusil}
\affiliation{Unit\'e Mixte de Physique CNRS/Thales, 1 Avenue A. Fresnel, 91767 Palaiseau, France and Universit\'e Paris-Sud, 91405 Orsay, France}
\author{K. Bouzehouane}
\affiliation{Unit\'e Mixte de Physique CNRS/Thales, 1 Avenue A. Fresnel, 91767 Palaiseau, France and Universit\'e Paris-Sud, 91405 Orsay, France}
\author{J.-M. Triscone}
\affiliation{D\'epartement de Physique de la Mati\`ere Condens\'ee, Universit\'e de Gen\`eve, 24 Quai Ernest Ansermet, 1211 Geneva 4, Switzerland}
\author{P. Paruch}
\affiliation{D\'epartement de Physique de la Mati\`ere Condens\'ee, Universit\'e de Gen\`eve, 24 Quai Ernest Ansermet, 1211 Geneva 4, Switzerland}
\date{\today}
\begin{abstract}
In studies using piezoresponse force microscopy, we observe a non-zero lateral piezoresponse at 180$^\circ$ domain walls in out-of-plane polarized, $c$-axis-oriented tetragonal ferroelectric Pb(Zr$_{0.2}$Ti$_{0.8}$)O$_3$ epitaxial thin films. We attribute these observations to a shear strain effect linked to the sign change of the $d_{33}$ piezoelectric coefficient through the domain wall, in agreement with theoretical predictions. We show that in monoclinically distorted tetragonal BiFeO$_3$ films, this effect is superimposed on the lateral piezoresponse due to actual in-plane polarization, and has to be taken into account in order to correctly interpret the ferroelectric domain configuration.\end{abstract}
\maketitle

Ferroelectric materials, characterized by their reversible spontaneous electric polarization, show great potential for multifunctional applications ranging from nonvolatile memories \cite{scott_memories,waser_memories} to nanoscale sensors and actuators \cite{damjanovic_sensors_review}. Controlling the structure and stability of ferroelectric domains in these materials is a key requirement for successful device implementation. In particular, the dynamics of domain walls, the interfaces separating regions with differently oriented ferroelectric polarization in the films, can significantly affect performance \cite{ouyang_ultrahigh_piezo}. Understanding domain wall behavior at the nanoscopic scales of current and future devices, however, requires techniques with the requisite nanoscale resolution.

One such technique is piezoresponse force microscopy (PFM) \cite{guthner_afm_FE}, in which  a metallic atomic force microscope (AFM) tip is used to apply an alternating voltage across the ferroelectric material, resulting in a local mechanical response at the film surface due to the inverse piezoelectric effect. This piezoelectric response can be detected from the induced displacement of the AFM  cantilever, recorded by the position of a laser beam reflected onto a quadrant-split photodetector. The vertical deflection and angular torsion of the tip are referred to as vertical and lateral PFM, respectively. The response phase provides information on the polarization, while its amplitude is related to the polarization magnitude \cite{kalinin_PFM_review}. Depending on the piezoelectric coefficient tensor $d_{ij}$, linked to the crystal symmetry, a combination of these two measurements allows access to both out-of-plane and in-plane components of polarization. Although quantitative measurements of piezoelectric coefficients via PFM are challenging,  the technique has provided valuable information about the behavior of domain walls and switching dynamics both in thin films \cite{tybell_creep,paruch_dw_roughness_FE} and in device structures \cite{gruverman_caps_afm, tiedke_AFM_caps,scott_nanodomain_faceting}.  In this context, understanding the origins of the PFM signal observed at ferroelectric domain walls is an important issue.

Considering only piezoelectric effects, in a $c$-axis-oriented tetragonal ferroelectric film with an electric field applied along the polarization axis, the piezoelectric response is determined by the $d_{33}$ coefficient, leading to a purely vertical PFM signal. However, in such films, a non-zero lateral PFM response has been observed specifically at the position of 180$^{\circ}$ domain walls. Initially, sliding \cite{wittborn_domain_imaging} or torque \cite{scrymgeour_pfm_180DW} mechanisms based on surface distortion as a result of antagonistic vertical contraction and expansion on either side of the domain wall were proposed. However, subsequent studies have questioned these scenarios after quantifiying the resulting lateral force,  and suggested instead the possibility of electrostatic effects arising from the electric field present around domain walls due to the build-up of opposite surface polarization charges \cite{jungk_DW_LPFM}. Concurrently, numerical analyses in the framework of resolution-function theory showed that in a  $c$-axis tetragonal film, local shear can occur at the domain wall \cite{morozovska_resolutionfunction_PFM},  potentially giving rise to a lateral PFM signal.  Discriminating between these different contributions to identify the  mechanism at the origin of the observed signal is all the more necessary as PFM becomes widely applied to more complex materials such as BiFeO$_3$ (BFO),  presenting both in-plane and out-of-plane polarization components, and where all possible contributions to both vertical and lateral PFM signals have to be taken into account.

In this letter, we report on local probe measurements of $c$-axis tetragonal Pb(Zr$_{0.2}$Ti$_{0.8}$)O$_3$ (PZT) thin films, which exhibit both the vertical PFM signal due to their polarization along the [001] axis, and a lateral PFM signal specifically at domain walls. Based on complementary electric force microscopy (EFM), we find the dominant contribution to the observed lateral signal to be piezoelectric, and due to shear at the domain walls. We compare these observations with similar measurements on monoclinically distorted tetragonal BFO, with polarization along the $<$111$>$ axes, and thus combined out-of-plane and in-plane polarization components with respect to the sample surface.

\begin{figure}[h]
\includegraphics[width=\columnwidth]{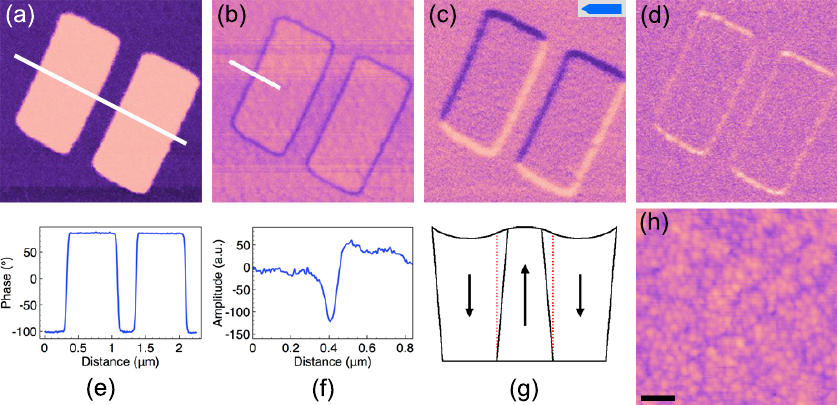}
\caption{PFM measurement of rectangular ferroelectric domains written on PZT by applying +12V to the scanning tip in an as-grown monodomain region (scale bar is 0.5 $\mu$m). Each domain structure shows a 180$^{\circ}$ contrast in the vertical phase (a),(e) with a corresponding minimal vertical amplitude at the domain wall (b),(f) (blue lines in (a) and (b) correspond to the profiles shown in (e) and (f)). Two opposite non-zero signals (dark and bright colors) are observed in the lateral phase (c) perpendicularly to the AFM cantilever at the position of these domain walls, with a corresponding rise in the lateral amplitude (d). The orientation of the AFM cantilever is indicated by the blue arrow in (c). (g) Schematical representation of domain wall shear deformation due to locally non-zero $d_{35}$ coefficient (dashed red lines show the initial domain wall positions). (h) Topography of the region, showing a 4 $\mbox{\AA}$ rms surface roughness.}
\label{pzt_pfm}
\end{figure}

70 nm PZT and BFO films on metallic SrRuO$_3$ (used as a bottom electrode during PFM measurements) were epitaxially grown on single-crystal (001) SrTiO$_3$ substrates by off-axis radio-frequency magnetron sputtering and pulsed laser deposition, respectively \cite{gariglio_pzt_films, bea_multiferroics_halfmetal_spintronics}. On these samples, we performed ambient-condition AFM measurements  with a \textit{Veeco Dimension V}. To create rectangular domains in the as-grown monodomain film, we applied a +12 V bias  to a scanning tip (\textit{MikroMasch} NSC18/Cr-Au). Subsequently, lateral and vertical PFM were simultaneously recorded for each domain structure, with ac bias voltage amplitude and frequency of 2-5 V and 20 kHz, respectively. The corresponding topographical images showed a typical rms surface roughness of 4 $\mbox{\AA}$ for PZT (\fref{pzt_pfm}h) and 2 nm for BFO.

For PZT films, as shown in \fref{pzt_pfm}a and e, we observe the expected 180$^\circ$ contrast between ``up'' and ``down'' polarized domains in the vertical PFM phase signal. Correspondingly, a minimum in the vertical PFM amplitude (\fref{pzt_pfm}b and f) is seen at the position of the 180$^\circ$ domain walls separating these regions. We also observe a clear non-zero lateral PFM phase signal, showing two opposite-contrast features  at domain walls (\fref{pzt_pfm}c), with corresponding maxima in the lateral PFM amplitude (\fref{pzt_pfm}d). 

The vertical PFM data demonstrates that between the ``up'' and ``down'' polarized regions  the $d_{33}$ coefficient of the material changes sign, locally going to zero at the 180$^\circ$ domain walls. In the shear strain scenario, the lateral PFM observations for purely tetragonal $c$-axis PZT can be explained by the presence of non-zero shear components ($d_{35}$) \cite{morozovska_resolutionfunction_PFM}. In uniformly polarized regions, as expected from symmetry considerations, the $d_{35}$ coefficient is zero, corresponding to no lateral PFM phase signal, and zero lateral PFM amplitude. However, at the domain walls, where the ferroelectric crystal symmetry breaks down, the local $d_{35}$ coefficient ($d_{35}^{DW}$) exhibits finite values, as reflected by the high lateral PFM amplitude signal. Moreover, the sign of the $d_{35}^{DW}$ response depends on the relative positions of the ``up'' and ``down'' domains with respect to the cantilever axis. In our measurements, two domain walls (``up'' on the left of the cantilever, ``down'' on its right, and vice-versa) are present on opposite sides of each domain, thus yielding two opposite-contrast features in the lateral PFM phase signal. We note that topographical contributions can be excluded due to low surface roughness, all the more so because the signals are clearly visible only at domain walls.

\begin{figure}[h]
\includegraphics[width= \columnwidth]{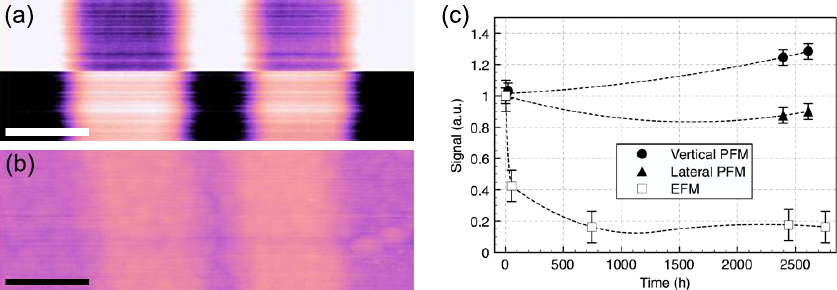}
\caption{EFM measurements of rectangular ferroelectric domains written on PZT, immediately (a) and three months (b) after writing, with same vertical scale of 5$^{\circ}$ (horizontal scale bars are 1.4 and 0.6 $\mu$m in (a) and (b), respectively). The upper and lower part of (a) correspond to +1 V and -1 V bias voltage applied to the tip during EFM, showing contrast inversion. (c) Time dependence of vertical and lateral PFM and EFM signals for domains written at $t=0$, showing a decay of the EFM signal vs. a stabilization of both PFM signals (dotted lines are a guide to the eye).}
\label{pzt_efm}
\end{figure}

However, the electrostatic contribution proposed by Jungk \textit{et al.} \cite{jungk_DW_LPFM} as the dominant mechanism could also produce similar effects. To discriminate between these two mechanisms, we performed complementary time dependence studies of EFM (in phase detection mode) vs. PFM signal measurements on ferroelectric domains (\fref{pzt_efm}). Previous similar studies on PZT thin films showed a decay of the EFM signal below noise levels after a few days, due to passivation by ambient charges and leakage \cite{tybell_switching_FE}. With both improved measurement setup and leakage properties in the PZT films used, we observed that an EFM signal is still present in the written domains three months after writing, although decreased by an order of magnitude from its initial value. However, both vertical and lateral PFM signals of these same domain walls do not show any significant decay over time, suggesting an effect of dominant piezoelectric nature.

In addition, we note that similar lateral PFM contrast has been observed in epitaxial PZT capacitors \cite{gruverman_caps_afm,gruverman_tutorial_aveiro}, which would effectively screen out the electrostatic contribution, again strongly suggesting that the dominant mechanism is piezoelectric. 

Taken together, our studies suggest that the mechanism behind the observed lateral PFM signal is indeed piezoelectric in nature, and due to the shear response resulting from a local variation of the tetragonal ferroelectric symmetry specifically at the 180$^\circ$ domain wall. This shear response of domain walls is also potentially of significant applied interest. In fact, a finite-element analysis of the  behavior of a piezoelectric transducer based on ferroelectric domain structures in PZT \cite{ballandras_SAW, ballandras_ppt_lno} suggests that such a lateral response, whose origin at that time remained unaddressed, could result in the generation of surface acoustic waves in the device \cite{kumar_SAW}.

Having considered the $d_{35}^{DW}$ response in materials with a  purely out-of-plane polarization, we next addressed its effects on PFM measurements in BFO thin films  deposited on (001) SrTiO$_3$, which present a monoclinically distorted tetragonal structure with polarization lying close to the $<$111$>$ directions \cite{zavaliche_BFO_domains,catalan_BFO_DW}. In these films, due to the lower symmetry and finite in-plane component of the polarization, the $d_{35}$ coefficient ($d_{35}^m$) is expected to be non-zero even in uniformly-polarized regions. To correctly characterize these films it is thus necessary to differentiate between any shear effect at the domain walls, and the actual in-plane polarization.

\begin{figure}[h]
\includegraphics[width=\columnwidth]{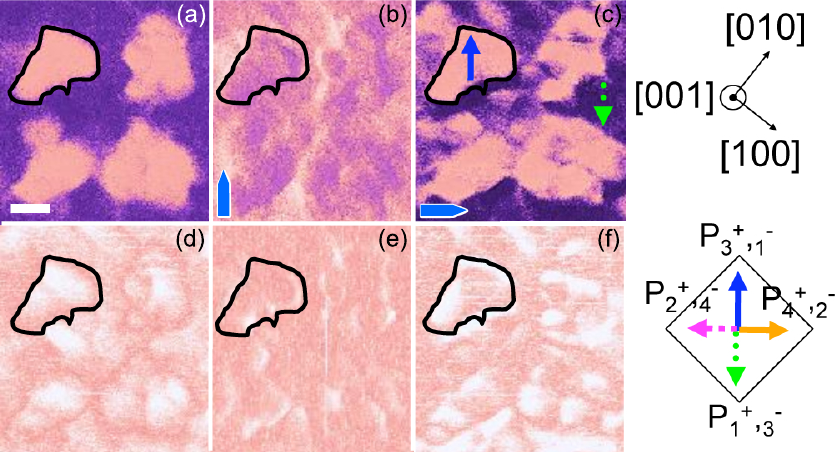}
\caption{PFM images of an array of domains written on BFO with +12V, 40 and 50 ms pulses on a prepolarized background: vertical (a), lateral (b),(c) PFM phases, vertical (d), lateral (e),(f) PFM amplitudes (horizontal bar is 100 nm). For the lateral measurements, the orientation of the cantilever is given by the blue arrows. The black contour in (a)-(f) corresponds to one of the written out-of-plane domain walls in (a). (g) Orientation of the crystalline axis of the substrate and the BFO film and of the in-plane projection of the eight possible orientations of the polarization. $+$ and $-$ stand for the out-of-plane component of the polarization going out or in the paper respectively.}
\label{Dimension_domains_BFO}
\end{figure}

In these BFO films, we performed PFM measurements for two perpendicular cantilever orientations, at 0$^\circ$ and 90$^\circ$ with respect to an array of small written domains (\fref{Dimension_domains_BFO}). As expected, the vertical PFM phase shows a 180$^\circ$ contrast between the written domains and the uniformly pre-polarized background (\fref{Dimension_domains_BFO}a), with the domain walls visible as a decrease of the vertical PFM amplitude (\fref{Dimension_domains_BFO}d). At the position of these written domains, we also observe changes in the lateral PFM signal. For a 0$^\circ$-oriented cantilever, the lateral PFM phase (\fref{Dimension_domains_BFO}b) is similar to the response observed in PZT, with two opposite-contrast features  at the position of the domain walls, and an otherwise zero response both in the uniformly polarized background and at the center of the written domains. An increased lateral PFM amplitude is observed at the corresponding domain walls (\fref{Dimension_domains_BFO}e), allowing the observed features to be attributed to the $d_{35}^{DW}$ response, as for PZT.  In the remainder of the image, the lack of a phase response suggests that only P$_{1,3}$$^{+,-}$ polarizations, as defined in \fref{Dimension_domains_BFO}g, are present in this region, with an in-plane projection parallel to the cantilever in this orientation, and thus not detectable by lateral PFM.

To confirm this observation, and to distinguish between the two different possible in-plane polarization orientations, we carried out lateral PFM measurements with the cantilever rotated by 90$^\circ$. In the phase images, as shown in \fref{Dimension_domains_BFO}c, a large bright region is now observed in the center of each domain, in contrast to the darker background. This phase contrast can be attributed to P$_1$$^-$ (full blue arrow) for the background and P$_1$$^+$ (dotted green arrow) for the domains, thus corresponding to 180$^\circ$ switching. Once again, at the domain walls, opposite-contrast features are present and superimposed on the $d_{35}^{m}$ response of the P$_1$$^-$ and P$_1$$^+$ polarizations. 
Finally, in the lateral PFM amplitude (\fref{Dimension_domains_BFO}f), we see a complex behavior, with a minimum at the position of the domain walls, as expected when the $d_{35}^{m}$ response changes sign. However, the amplitude signal appears to increase in the immediate vicinity of the domain walls. This behavior can be interpreted as a combination of the $d_{35}^{DW}$ response at the domain wall and of the  $d_{35}^m$ response of the P$_1$$^-$ and P$_1$$^+$ domains.

In conclusion, we have shown using a combination of PFM and EFM measurements, that the mechanism behind the lateral PFM response observed at 180$^{\circ}$ domain walls in purely out-of-plane polarized tetragonal PZT thin films is indeed piezoelectric, and related to the shear response specifically at such domain walls. This effect could be useful for nanomechanical transducers in which horizontal propagation of a surface deformation is required. The lateral shear at domain walls should be treated with great care when interpreting lateral PFM images, especially when dealing with ferroelectrics that present an actual in-plane component of the polarization such as BFO.  More complex effects could also be expected in PFM measurements on materials with a polycrystalline granular morphology.  For a more quantitative interpretation of such images, calculations of the $d_{35}^{DW}$ response in materials such as BFO would be extremely useful. 

\begin{acknowledgments}
The authors thank M. Bibes, A. Barth\'el\'emy, E. Soergel, and T. Jungk for helpful discussions, and M. Lopes for technical support. This work was supported by the Swiss National Science Foundation through the NCCR MaNEP and Division II, and by the European Commission STREP project MaCoMuFi. H. B. is supported by Bourse d'excellence from the University of Geneva.
\end{acknowledgments}

\newpage

\bibliographystyle{prsty}

\end{document}